\documentclass[11pt,preprint]{aastex}

\begin{document}

\title{WHITE DWARF COSMOCHRONOLOGY IN THE SOLAR NEIGHBORHOOD}

\author{P.-E. Tremblay$^{1,2}$, J.~S. Kalirai$^{1,3}$, D.~R. Soderblom$^{1}$, M. Cignoni$^{1}$, and J. Cummings$^{3}$}

\affil{$^{1}$Space Telescope Science Institute, 700 San Martin Drive,
  Baltimore, MD 21218} \affil{$^{2}$Hubble Fellow} \affil{$^{3}$Center
  for Astrophysical Sciences, Johns Hopkins University, 3400 North
  Charles Street, Baltimore, MD 21218} \email{tremblay@stsci.edu}

\begin{abstract}

The study of the stellar formation history in the solar neighborhood
is a powerful technique to recover information about the early stages
and evolution of the Milky Way. We present a new method which consists
of directly probing the formation history from the nearby stellar
remnants. We rely on the volume complete sample of white dwarfs within
20~pc, where accurate cooling ages and masses have been
determined. The well characterized initial-final mass relation is
employed in order to recover the initial masses (1 $\lesssim M_{\rm
  initial}/M_{\odot} \lesssim$ 8) and total ages for the local
degenerate sample.  We correct for moderate biases that are necessary
to transform our results to a global stellar formation rate, which can
be compared to similar studies based on the properties of
main-sequence stars in the solar neighborhood. Our method provides
precise formation rates for all ages except in very recent times, and
the results suggest an enhanced formation rate for the solar
neighborhood in the last 5~Gyr compared to the range 5 $<$ Age (Gyr)
$<$ 10. Furthermore, the observed total age of $\sim$10~Gyr for the
oldest white dwarfs in the local sample is consistent with the early
seminal studies that have determined the age of the Galactic disk from
stellar remnants. The main shortcoming of our study is the small size
of the local white dwarf sample. However, the presented technique can
be applied to larger samples in the future.

\end{abstract}

\keywords{white dwarfs -- Galaxy: disk -- Galaxy: stellar content --
  Galaxy: evolution -- solar neighborhood}

\section{INTRODUCTION}

The detailed study of stars in the solar neighborhood allows for the
calibration of stellar structure and evolution models. This can be
done for instance with precise measurements of the effective
temperature, luminosity, and metal abundance of local stars in order
to compare with predicted isochrones, or by surveying binaries where
the age and distance of all components is expected to be
identical. One significant advantage of the local sample is the
abundance of data and the feasibility of creating large volume
complete samples.  It is then possible to learn about the stellar
formation history (SFH) and the initial mass function (IMF), in
principle for different Galactic components, i.e. the thin disk, thick
disk, and halo, if one is able to identify independently the
populations from kinematics or metallicities. Various studies have
been aimed at identifying the SFH of the disk, with different
techniques, and often with conflicting results. The most common
approach has been to invert the observed color-magnitude diagram into
a SFH using stellar isochrones \citep{hernandez00,SFR1,SFR2}. Another
group of studies have been employing stellar activity in low-mass
stars as an indicator of age \citep{soderblom91,rocha00,fuchs09}.
Other techniques to derive the SFH include empirical age versus
metallicity relations \citep[see, e.g.,][]{reid07} and the age
distribution of nearby open clusters \citep{fuente04}.

Age determination for individual stars is difficult with any method
\citep[see the review of][]{soderblom10}, especially for large ages
where the stellar activity is small, the age versus metallicity
relation is uncertain, and the age sensitivity of the color-magnitude
diagram is greatly diminished except for a few evolved stars.  Even
for the SFH in the last 5~Gyr, studies find rates ranging from a
nearly constant value \citep{rocha00,reid07} to a significantly peaked
distribution with a maximum 3-5~Gyr ago \citep{SFR1,SFR2,fuchs09}. In
this work, we rely instead on local stellar remnants, which can be studied
to derive very accurate masses and cooling ages.

The white dwarf luminosity function, defined as the number of stars as
a function of their intrinsic luminosity, has been used as a tool to
determine the age of the thin disk
\citep{winget87,liebert88,leggett98}. Further studies have also
revealed the white dwarf spatial density and integrated formation
rates \citep{liebert05,harris06,limoges10}. The luminosity function
can also, in principle, provide information about the stellar
formation history \citep{diaz-pinto94}, although it is difficult to
extract this quantity because of the degeneracy between mass and age
at constant luminosity. \citet{SFRWD} have demonstrated the
possibility of recovering the stellar formation history from an
inversion of the luminosity function. Nevertheless, the observed
luminosity functions are drawn from samples that are not volume
complete, such as the Sloan Digital Sky Survey \citep{harris06}, and
complex corrections for completeness and contaminations have to be
made.

On a different front, several works have been aimed at identifying a
complete volume-limited sample of white dwarfs around the
Sun. \citet{holberg02} was the first study dedicated to this sample,
where they estimated at the time that within 20~pc, only 65\% of the
stellar remnants were known. The goal of achieving a complete volume
sample was pursued by different studies \citep{holberg08,sion09}, and
\citet{giammichele12} most recently presented a homogeneous review of
the 20~pc sample with a consistent set of model atmospheres in order
to improve the derived stellar parameters and distances. They examine
the properties of 168 potentially close white dwarfs, and by comparing
the space number density of the 13~pc and 20~pc samples, they estimate
that the latter is more than 90\% complete.

In this work, we rely on the results of \citet[][hereafter
  GB12]{giammichele12} to study the SFH in the local neighborhood. The
significant advantage of this sample is that the white dwarfs have
precise distances, luminosities, masses, and cooling ages. This allows
for a direct conversion of the remnant parameters to initial stellar
parameters, employing the well studied initial-final mass relation
calibrated from clusters and binaries
\citep{kalirai05,catalan08,kalirai08,kalirai09,williams09,dobbie12},
and stellar isochrones for the main-sequence lifetime. While there are
still biases in the derivation of the global stellar formation history
since not all stars have become white dwarfs, our technique does not
involve the calculation of a luminosity function where some of the
information is lost. We compare our results to previous studies,
including those relying on main-sequence stars. The local white dwarf
sample is still fairly small with only around one hundred objects,
however the proposed technique can be used in future studies. For
instance, {\it Gaia} is expected to identify a volume complete sample
of degenerates up to $\sim$40~pc, including accurate individual distances
and masses from parallaxes and photometry, and stellar population
identifications from proper motions \citep{carrasco14}. In Section~2,
we review the observed degenerate star sample. We follow in Section~3
with our derived SFH in the solar neighborhood. In Section~4, we
characterize our uncertainties and compare our results to those
obtained with other independent techniques and observations. The
conclusion follows in Section~5.

\section{WHITE DWARF SAMPLE}

We rely directly on the white dwarf parameters and associated
uncertainties derived in Table~2 of GB12. The atmospheric parameters
were determined from a combination of photometric, spectroscopic, and
parallax observations, and we refer to GB12 for a complete description
of the sample and observations. In most cases, the photometric
technique provided the best constraint on the fundamental parameters,
where the combination of the photometric fluxes and parallaxes allowed
for $T_{\rm eff}$, radius, and distance determinations.  The total
mass and cooling age were then derived employing the evolutionary
models of \citet{fontaine01}. These models have C/O cores (50/50 by
mass fraction mixed uniformly) and assume thick hydrogen layers
($M_{\rm H}/M_{\rm total} = 10^{-4}$) for H-atmosphere white dwarfs
and thin layers ($M_{\rm H}/M_{\rm total} = 10^{-10}$) for helium and
mixed atmospheres. When possible, the Balmer lines in the spectra were
also compared with model atmospheres to provide both $T_{\rm eff}$ and
$\log g$ \citep{bergeron92}. The evolutionary models described above
were then used to determine the radius, mass, luminosity, cooling age,
and finally distance in combination with the observed magnitude.

The atmospheric parameters were derived using model atmospheres from
\citet{tremblay11}, \citet{bergeron11}, and \citet{dufour05,dufour07},
for pure-hydrogen, pure-helium, and metal-rich (DQ, DZ) atmospheres,
respectively. The pure-hydrogen atmospheres include the Ly-$\alpha$
red wing opacity \citep{kowalski06}. In the meantime, a new grid of
predicted spectra for pure-hydrogen atmospheres including a treatment
in 3D of the convection has been published
\citep{tremblay13}. However, GB12 (see Figure 16) have already included
a correction to their spectroscopically determined atmospheric
parameters based on early results from the 3D simulations
\citep{tremblay11}, and suggested an empirical correction assuming
that the high-mass bump seen in the larger SDSS sample is entirely due
to shortcomings in the 1D model atmospheres. The latest 3D corrections
from \citet{tremblay13} are fairly similar to the empirical correction
used in GB12, despite the fact that they did not include a $T_{\rm
  eff}$ correction. Since we bin the data in 1~Gyr intervals in this
work, it is a very good approximation to use directly the results of
GB12.

\begin{figure}[p]
\epsscale{0.8}
\plotone{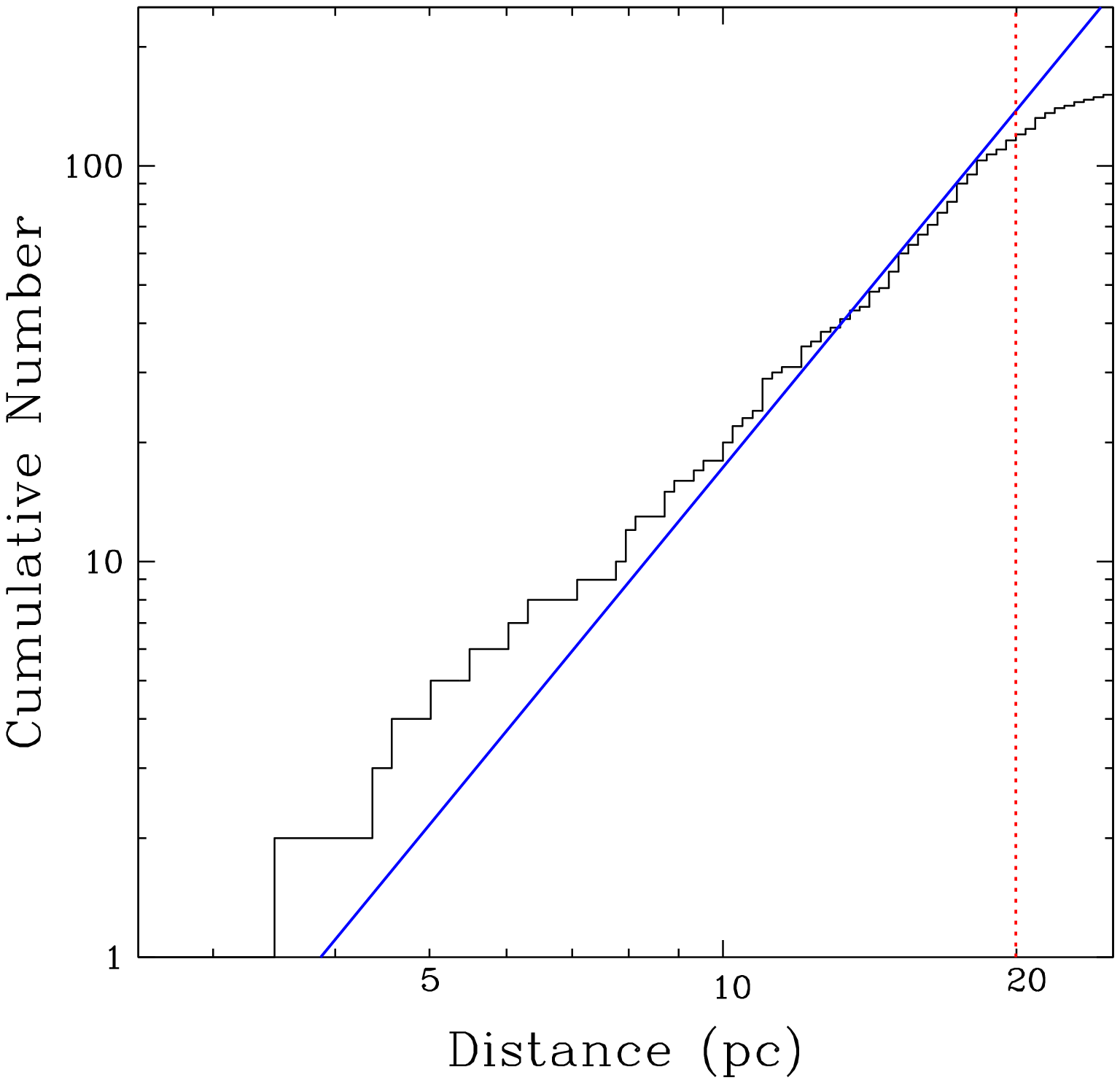}
\begin{flushright}
\caption{Number of white dwarfs as a function of distance (logarithm
  scales) for the local sample of GB12. The solid blue line represents
  the expected uniform distribution of stars for a volume complete
  sample normalized at the number of objects at 13~pc. The vertical
  dotted red line represents the 20~pc limit of this
  work. \label{fg:f_complete}}
\end{flushright}
\end{figure}

In the following, we only keep objects with a derived distance smaller
than 20 pc, for a total of 117 remnants. The sample does not include
three white dwarfs in Sirius-like systems \citep{holberg13} where the
data on the degenerate counterpart is insufficient to derive the
atmospheric parameters.  Figure~\ref{fg:f_complete} reviews the 20~pc
sample completeness by showing the cumulative number of objects as a
function of distance. We compare with the expected +3 log-log slope
for a complete sample, normalized at 13~pc assuming the sample is
complete at this distance \citep{holberg08}.  According to this
normalization, the 20~pc sample is only 82\% complete in contrast to
the 90\% value quoted in GB12. Their higher estimate is mostly because
they increased the sample size by including objects that could lie
within the uncertainties inside the 20~pc region, and also because
they relied on the 13~pc number density from \citet{holberg08}. The
completeness already reaches 92\% at 18~pc according to our results,
hence we can review the integrity of the sample using both the 18 and
20~pc boundaries.

\section{STELLAR FORMATION HISTORY}

In order to study the SFH in the solar neighborhood, we need to
recover the initial stellar parameters from the remnant
properties. White dwarfs have been observed in different populations
at solar metallicity to derive initial-final mass relations that are
in relatively good agreement. We rely on the prescription of
\citet{kalirai08} who studied in particular the low-mass end of the
initial-final mass relation in two open clusters\footnote{We also use
  the more recent globular cluster constraints from
  \citet{kalirai09}.}. The low-mass end is critical to study old stars
in the solar neighborhood, and difficult to observe due to the lack of
close old clusters. The Kalirai et al. relation covers final masses in
the range $0.53 \leq M_{\rm final}/M_{\odot} \leq 1.02$ corresponding
to $0.8 \leq M_{\rm initial}/M_{\odot} \leq 6.5$. We use a third-order
polynomial fit to the data.
 
We employ directly the white dwarf cooling ages derived in GB12 from
the evolutionary models of \citet{fontaine01}. The total age is the
sum of the cooling age and the main-sequence lifetime from the
evolution calculations of \citet{hurley00}, assuming a solar
metallicity, and the initial mass derived from the initial-final mass
relation discussed above. Figure~\ref{fg:f_dist} presents the initial
mass as a function of total age (lookback time) for the white dwarfs
in the 20~pc sample. For 12 objects (red circles on the figure), no
mass information is available, hence we assume the canonical $\log g =
8.0$ value to determine the initial stellar parameters. There are 13
objects, with a derived mean mass of 0.44 $\pm$ 0.06 $M_{\odot}$, for
which the total age would be significantly larger than the age of the
universe. These objects are likely unresolved binaries (see GB12 and
\citealt{brown11}) and we exclude them from our analysis.

\subsection{Biases}

Figure~\ref{fg:f_age} presents the number of white dwarfs in 1~Gyr age
bins directly from the initial mass versus age distribution of
Figure~\ref{fg:f_dist}, which we define as the {\it raw} SFH (dashed
red line). We also display in Figure~\ref{fg:f_age} our best SFH
estimate (filled black histogram) considering observational biases
that we describe in the following sections. To begin, the total SFH is
the sum of objects that are at present day white dwarfs, stars, and in
much smaller number giants.  In Section~3.1.1 we correct for the {\it
  missing main-sequence star bias}. The 20~pc sample is close to the
central plane of the Galactic disk and populations with a small
velocity dispersion in the vertical Galactic coordinate are over
represented. We correct for this {\it kinematic bias} in
Sections.~3.1.2. Finally, Section~4.1 is devoted to other possible
biases in order to highlight the uncertainties of our derivation.

\subsubsection{Main-Sequence}

\begin{figure}[p]
\epsscale{0.8}
\plotone{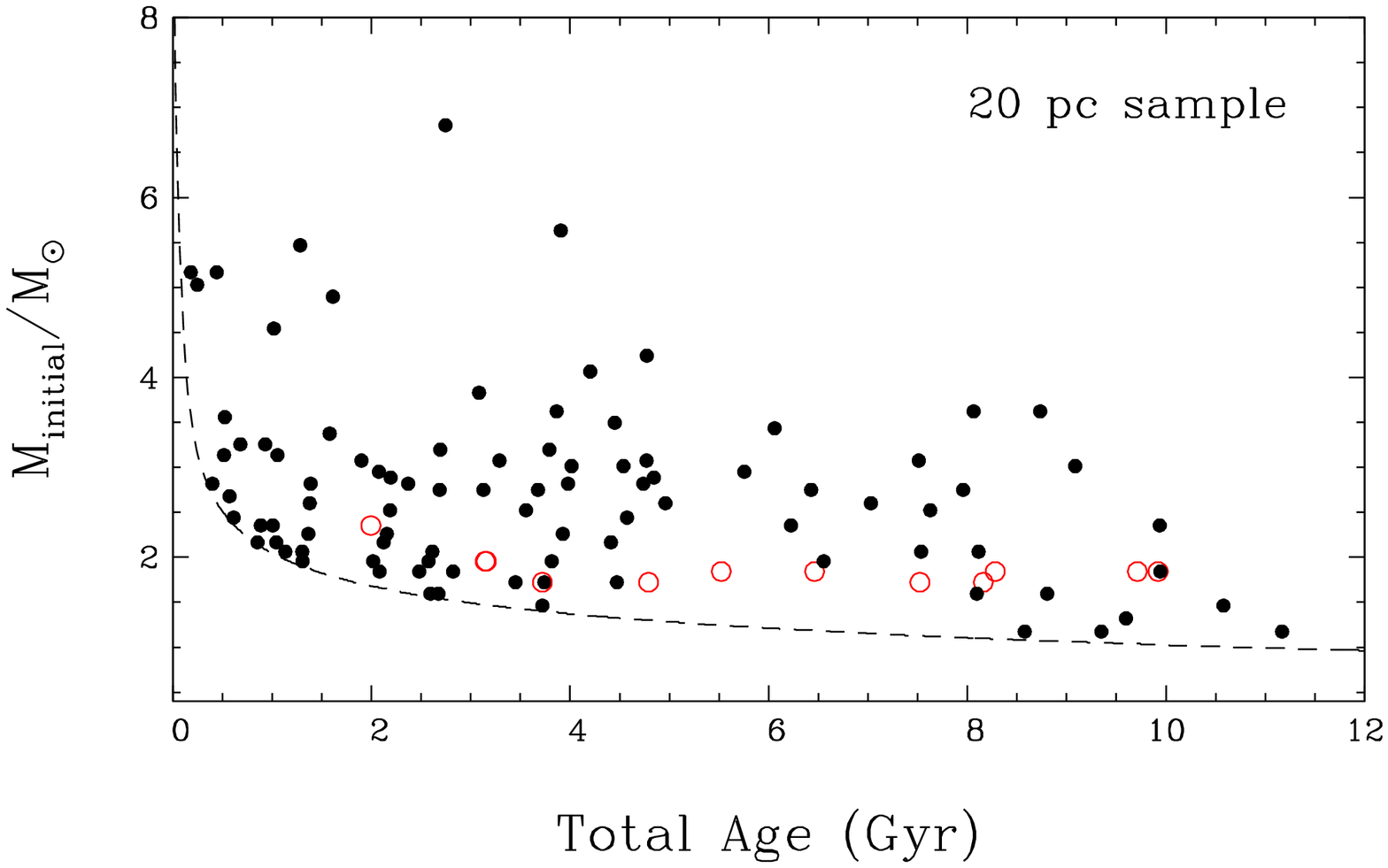}
\begin{flushright}
\caption{Initial mass distribution for the white dwarfs in the local
  20~pc sample of GB12 as a function of total age. Masses were derived
  with the initial-final mass relation of \citet{kalirai08} and the total
  age is the sum of the white dwarf cooling time \citep{fontaine01}
  and the main-sequence lifetime \citep{hurley00}. The dashed curve
  identifies where the total age is equal to the main-sequence
  lifetime, hence below which white dwarfs have not formed yet at
  present-day. The remnants with a fixed $\log g = 8.0$ value are
  identified with open red circles. \label{fg:f_dist}}
\end{flushright}
\end{figure}

\begin{figure}[p]
\epsscale{0.8}
\plotone{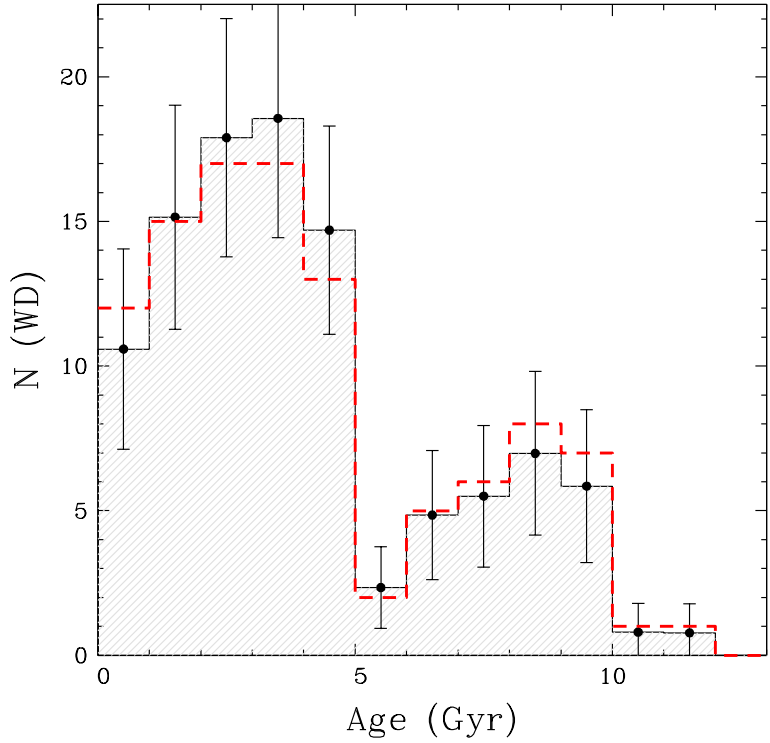}
\begin{flushright}
\caption{Number of white dwarfs in 1~Gyr total age bins (red dashed curve) 
  from the data of Figure~\ref{fg:f_dist}. The
  black filled histogram takes into account the biases due to the
  missing main-sequence stars (see Sect.~3.1.1) and the velocity
  dispersion $\sigma_{\rm W}$ in the Galactic coordinate $W$ (see
  Sect.~3.1.2), and has been normalized for the same total number of
  stars. The error bars take into account number statistics
  uncertainties and are derived from the uncorrected number of white
  dwarfs. \label{fg:f_age}}
\end{flushright}
\end{figure}

The main-sequence lifetime is larger than the lookback time in the
initial mass versus age area below the dotted line in
Figure~\ref{fg:f_dist}. This region would be populated with H-burning
stars that are excluded from our sample. In order to derive the total
SFH in the Galactic disk, we need to account for both main-sequence
and degenerate stars. Calculating an absolute formation rate would be
very difficult from white dwarfs alone since most of the stars are
formed as M dwarfs, which are still on the main-sequence. In order to
derive the relative formation rate, it is unnecessary to count all
local stars, but we still have to account for the change, as a
function of lookback time, of the ratio in number between white dwarfs
and main-sequence stars in the present day population.

Figure~\ref{fg:f_dist} demonstrates that the separation between the
white dwarf and stellar content of the Galaxy is fairly similar as a
function of age. This is ensured by the rapid variation of the
main-sequence lifetime as a function of initial mass. One exception is
for the last billion year bin where only stars born much more massive
than the Sun became white dwarfs. To evaluate this bias, we computed
the ratio of stars, assuming a \citet{salpeter} IMF, below and above
the threshold between a present day white dwarf and a main-sequence
star (dashed line in Figure~\ref{fg:f_dist}). This ratio $B_{\rm
  MS-WD}$ is defined as

\begin{equation}
B_{\rm MS-WD} = \frac{\int_{M_0}^{M_{\rm lim}} M^{-2.35} dM
}{\int_{M_{\rm lim}}^{M_1} M^{-2.35} dM} ~,
\end{equation}

{\noindent}where

\begin{equation}
M_{\rm lim} = M(t_{\rm lookback}=t_{\rm main-sequence~lifetime}) ~,
\end{equation}
  
\begin{figure}[p]
\epsscale{0.8}
\plotone{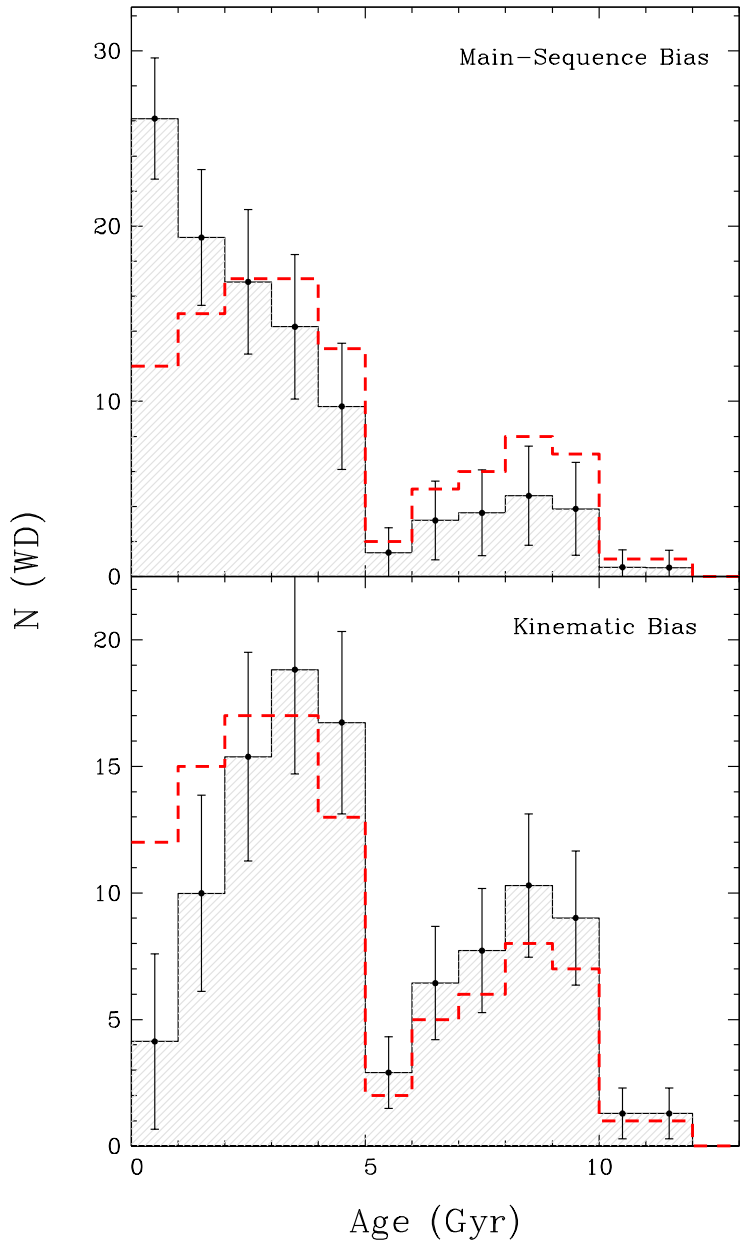}
\begin{flushright}
\caption{Similar to Figure~\ref{fg:f_age} with the raw data from
  Figure~\ref{fg:f_dist} shown with a red dashed curve. The black filled
  histograms account for only one bias, due to missing main-sequence
  stars (top panel), and velocity dispersion (bottom panel),
  respectively. The distributions were re-normalized for the same total number of
  objects. The error bars take into account number statistics and
  are derived from the uncorrected number of white dwarfs.
\label{fg:f_bias12}}
\end{flushright}
\end{figure}

{\noindent}while $M_{0}$ and $M_{1}$ are some arbitrary small and
large masses outside of the $M_{\rm lim}$ range surveyed by our
study. Values of $M_{0}$ and $M_{1}$ do not matter since we are only
interested in the relative SFH. The obtained correction $B_{\rm
  MS-WD}$ is simply multiplied by the number of stars in each age bin
and the full distribution is then re-normalized to the actual number
of white dwarfs. Figure~\ref{fg:f_bias12} (top panel) shows the effect
of this bias alone in comparison to the raw distribution. The
strongest effect is for ages below 1~Gyr, where the steepness of the
IMF allows for few white dwarfs to be formed compared to larger ages.

In principle, it would be possible to compute the average IMF,
i.e. integrated over all ages, directly from the results of the local
white dwarf sample in Figure~\ref{fg:f_dist}. However, the small size
of the sample and properties of the IMF imply that only a small number
of objects have masses $1\sigma$ higher than the average, and it is
likely that the $\sim$15\% missing white dwarfs in the sample are
fainter hence more massive than the average. GB12 also suggest that
the most massive objects may be the result of mergers. Therefore, we
refrain from using the observed IMF, although we note that it is
consistent with a fairly bottom-heavy IMF, with a power-law as much as
1 dex steeper than the Salpeter relation. This result is compatible
with the lack of massive white dwarfs in the Hyades, just outside the
20~pc sample, considering a Salpeter IMF
\citep{williams04,tremblay12}. On the other hand, it is at odds with
the Salpeter-like relation observed, on average, in nearby clusters
\citep[see, e.g.,][]{imf-dan}.  We hope that larger and more complete
white dwarf samples will be able to use the observed IMF in a
consistent way to derive the SFH.

\subsubsection{Kinematics}

\begin{figure}[p]
\epsscale{0.8}
\plotone{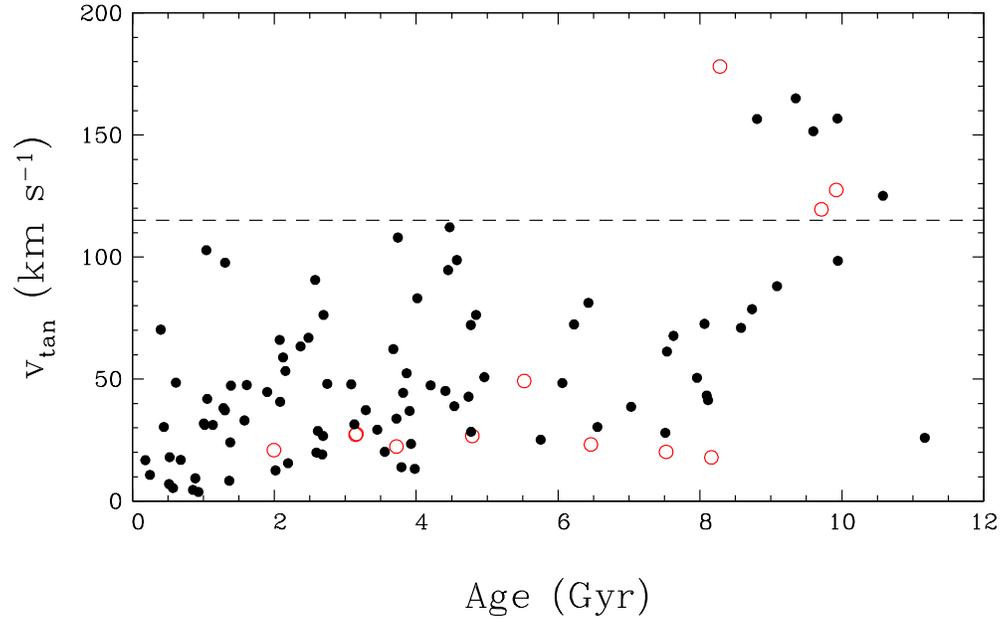}
\begin{flushright}
\caption{Tangential velocities for the white dwarfs in the local
  sample as a function of total age. $v_{\rm tan}$ was computed from
  the known distances (GB12) and proper motions
  \citep{sion14}. Remnants with a fixed $\log g = 8.0$ value are
  identified with open red circles.\label{fg:f_vel}}
\end{flushright}
\end{figure}

In this section we correct the SFH for kinematics. We have compiled
the proper motions for the local sample \citep{sion09,sion14} and
Figure~\ref{fg:f_vel} demonstrates a tangential velocity ($v_{\rm
  tan}$) versus age relation, where the velocity was computed using
distances from GB12. While most objects are consistent with thin disk
kinematics, it is difficult to disentangle thin versus thick disk
populations even based on 3D kinematics \citep{kawka06,sion09}.  We
note, however, the presence of a correlation between age and velocity,
and in particular a significant $v_{\rm tan}$ increase in the range $9
< {\rm Age~(Gyr)} < 11$, a population which represents about $8\%$ of
the sample. The derived total ages in this range are sensitive to the
parameterization of the initial-final mass relation (see
Sect.~4.1). Furthermore, three objects do not even have mass
  measurements. These objects are relatively old with a mean cooling
  age of 7.8~Gyr, assuming $\log g = 8.0$, hence they could be the
  remnant of short-lived main-sequence stars, be closer, and have
  smaller velocities than assumed. Therefore, it is difficult to
confirm the mean age of this population, and whether it represents the
tail of the thin disk or a separate thick disk
population. \citet{sion09} argue that there is no obvious separation
between the thin and thick disk populations in the 20~pc sample, an
interpretation that is in agreement with the review of the stellar
content of the SDSS SEGUE survey showing no distinct thick disk
component in our Galaxy \citep{bovy12}. On the other hand, the UVW 3D
space motions of the white dwarfs in the SN Ia Progenitor survey
\citep[SPY;][]{pauli06} revealed that 7\% of their sample belongs to
the thick disk, in agreement with the local sample at face
value. \citet{reid05} suggests a thick disk population of as much as
20\% in the solar neighborhood, which could still be compatible with
the local sample considering that high-velocity components are under
represented in a volume-complete sample.

The population identification for the local sample is not essential
for our study as long as one keeps in mind that the derived SFH might
not exclusively account for the thin disk. More critical is the bias
caused by changes in the dispersion of the vertical component of the
velocity in Galactic coordinates ($\sigma_{\rm W}$) as a function of
age. We refrain from studying this issue directly from the local white
dwarfs since the subsample having 3D velocities is far from complete
and not homogeneous. Instead we rely on the studies of
\citet{nordstrom04} and \citet{seabroke07} who determined $\sigma_{\rm
  W}$ versus age for a large sample of nearby F and G stars. Both
studies agree that $\sigma_{\rm W}$ increases by a factor of two in
the 1-5~Gyr age range. \citet{seabroke07} propose a new binning
procedure and suggest the relation

\begin{equation}
\sigma_{\rm W} = k~{\rm age}^{0.6} ~,
\end{equation}
  
{\noindent}where $k$ is a constant. They demonstrate that it is
unclear whether this trend continues for ages larger than 5~Gyr or if
there is a saturation at constant $\sigma_{\rm W}$ for thin disk
stars. We consider the latter possibility as our standard correction,
but review the former possibility in Section~4.1.  The volume bias
correction is at first order directly proportional to $\sigma_{\rm W}$
and is shown in Figure~\ref{fg:f_bias12} (bottom panel) compared to
the raw data. This correction has a slope, as a function of age, with an
opposite sign compared to the missing main-sequence star bias described
in the previous section, hence the effects largely cancel each other out 
in our resulting best estimate of the SFH in Figure~\ref{fg:f_age}.

\section{DISCUSSION}

\subsection{Assessing Biases}

\begin{figure}[p]
\epsscale{0.8}
\plotone{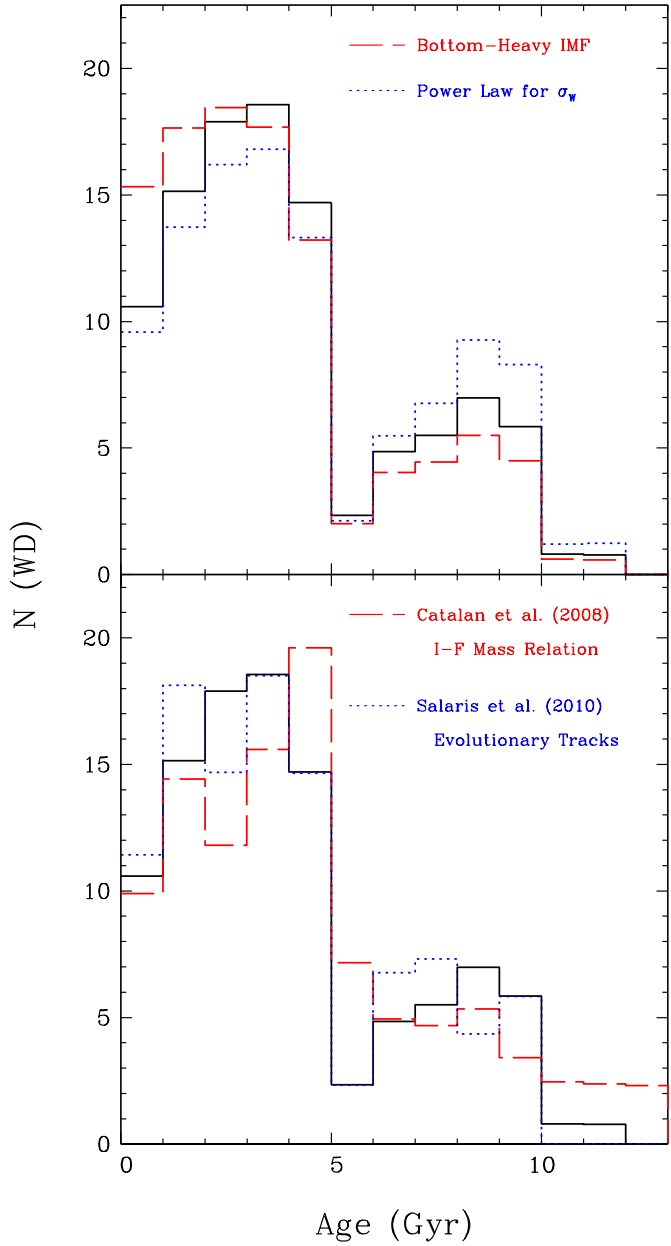}
\begin{flushright}
\caption{Alternative derivations of the SFH compared to our best
  standard estimate of Figure~\ref{fg:f_age} (solid black curves). All
  distributions were re-normalized to have the same total number of
  objects. {\it Top panel:} We rely on a steeper power law IMF with an
  index of $-3.2$ instead of $-2.35$ for the standard Salpeter case
  (dashed red). We use the power law of Eq.~3 for the $\sigma_{\rm W}$
  vs. age relation at all ages (dotted blue) instead of assuming a
  saturation above 5~Gyr. {\it Bottom panel:} We employ the
  initial-final mass relation of \citet[][dashed red]{catalan08}
  instead of the one from \citet{kalirai08}. Finally, we take the
  \citet{salaris10} cooling tracks (dotted blue) instead of those from
  \citet{fontaine01}.
 \label{fg:f_biases1}}
\end{flushright}
\end{figure}

\begin{figure}[p]
\epsscale{0.8}
\plotone{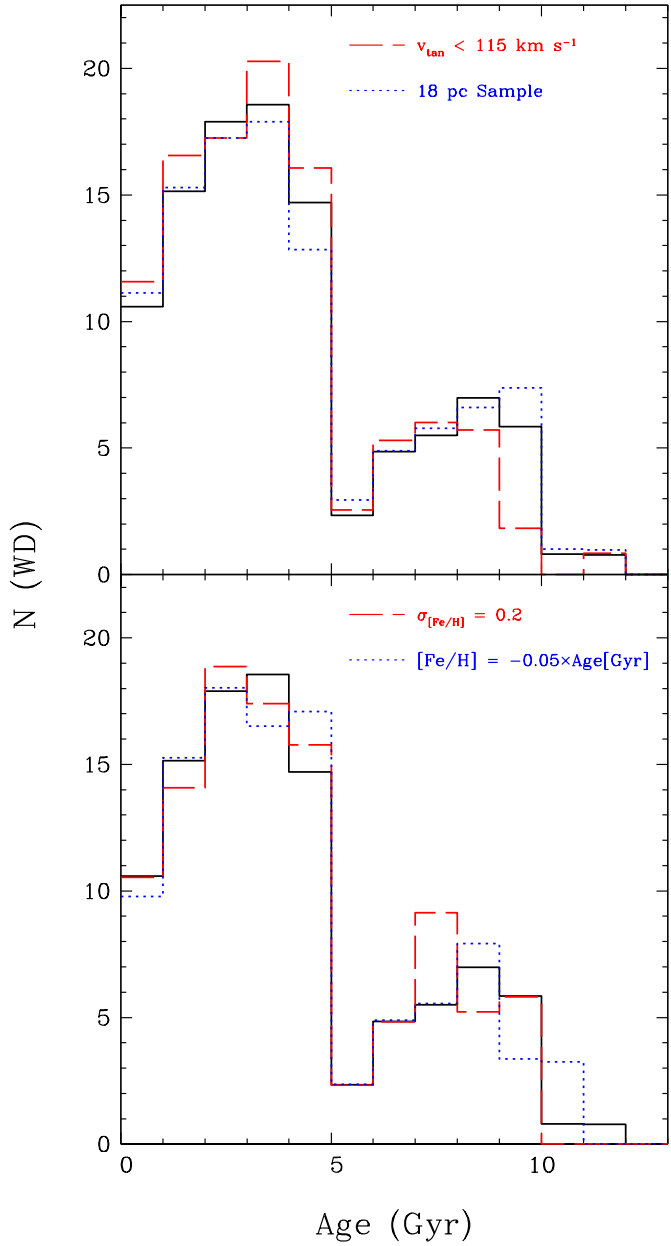}
\begin{flushright}
\caption{Similar to Figure~\ref{fg:f_biases1} but with additional
  alternative derivations of the SFH. All distributions are
  re-normalized to the same integrated number of stars. {\it Top
    panel:} The high velocity population ($v_{\rm tan} >$
  115~km~s$^{-1}$) is removed (dashed red), and the more complete
  18~pc sample is used instead of the standard 20~pc sample (dotted
  blue). {\it Bottom panel:} The main-sequence lifetimes are computed
  with metallicities derived from a Monte-Carlo simulation with a
  dispersion of $\sigma_{\rm [Fe/H]}$ = 0.2 around solar metallicity
  (dashed red), or a linear age vs. metallicity relation with a solar
  value at present time and a value of [Fe/H] = $-$0.5 at 10 Gyr
  (dotted blue). \label{fg:f_biases2}}
\end{flushright}
\end{figure}

We review in turn the different biases and uncertainties in our best
estimate of the SFH presented in Figure~\ref{fg:f_age}. Several
experiments are presented in Figures~\ref{fg:f_biases1}
and~\ref{fg:f_biases2} and described in this section.  First of all,
in Figure~\ref{fg:f_biases1}, we use a steeper theoretical power-law
for the IMF, with an index of $-3.2$ instead of the commonly used
Salpeter law with a slope of $-2.35$ as a function of increasing $\log
M_{\odot}$. The steeper IMF is closer to the observed value for the
local sample. The impact is mostly seen for the first age bin, where
the bias correction for the ratio between white dwarfs and stars is
critical. This demonstrates that due to the uncertain IMF, we can not
constrain very well the slope of the SFH in the last 3~Gyr, although
this does not change much the overall shape of the SFH.  We have also
used an alternative description of the $\sigma_{\rm W}$ vs. age
relation, by relying on Eq.~3 at all ages, which is more consistent
with the interpretation of \citet{nordstrom04}. It increases the bias
correction for old white dwarfs in Figure~\ref{fg:f_biases1}, although
the overall shape of the SFH is similar.

The following experiment in Figure~\ref{fg:f_biases1} (bottom panel)
employs the \citet{catalan08} initial-final mass relation instead of
that of \citet{kalirai08}. These studies were the first to put
significant constraints on the low-mass end of the relation, the
former by examining white dwarfs in common proper motion pairs. The
results demonstrate that total age uncertainties are of the order of
1~Gyr since a significant amount of white dwarfs are shifted to the
next bin. There are more objects with a total age older
than 10~Gyr when using the Catal{\'a}n et al. relation. This confirms the
difficulty of assigning a population membership to the oldest white
dwarfs in the local sample, although the overall shape of the SFH does
not depend appreciably on the initial-final mass relation.  Finally,
the last experiment in Figure~\ref{fg:f_biases1} takes the alternative
white dwarf cooling sequences of \citet{salaris10} as input for the
total age. We rely on the sequences including the effects of C/O
  phase separation and sedimentation in the core. Since radii were
not available in their table, we still used the mass-radius relation
of \citet{fontaine01}. Nevertheless, it illustrates that differences
in the independent cooling models do not have a significant impact on
our results.

The second series of experiments in Figure~\ref{fg:f_biases2} starts
with a tangential velocity cutoff in order to remove a population that
is potentially not part of the thin disk. We made the cutoff at
$v_{\rm tan} < 115$~km~s$^{-1}$ based on the results of
Figure~\ref{fg:f_vel}. While there is a more rapid dip in the SFH for
ages larger than 9~Gyr, the overall shape at earlier ages is
unchanged. Another uncertainty comes from the incompleteness of the
sample as well as the incomplete information for 12 objects with no
$\log g$ determinations. Since the latter objects are mostly cool
white dwarfs with no or weak spectral features, they are unambiguously old and
it is unlikely that $\log g$ determinations, even if the mean value
was significantly different to 8.0, would change the SFH picture. We
estimated in Section~2 that the sample is 82\% complete. To illustrate
the impact of this bias, we rely instead on the 18~pc sample, which is
expected to be 92\% complete, to derive the SFH in
Figure~\ref{fg:f_biases2}. The effect is relatively mild, and there is
no clear age dependence as one could have expected since missing white
dwarfs are more likely to be fainter than the average.  However, this
does not necessarily imply very old ages for the faintest remnants,
since a white dwarf with a 2~Gyr cooling time already has a
temperature of $\sim$6000 K, which is not far from the coolest objects
in the sample ($T_{\rm eff} \sim 4000$~K).

The last uncertainty that we discuss in this section is the assumption
that all stars were formed with a solar metallicity.  This is based on
the fact that we can not recover the initial metallicity from white
dwarf observations.  The age vs. metallicity relation is not very well
understood \citep{soderblom10}. It impacts first of all the total
main-sequence lifetime, and there are indications it may also impact
the initial-final mass relations \citep{zhao12}, although we already
accounted for this uncertainty above by looking at two different
relations calibrated from different populations (and metallicities).
We made a Monte Carlo experiment in which the initial metallicities
for the local sample were randomly selected from a normal distribution
with a dispersion of [Fe/H] = 0.2 and a mean solar value. This
supposition corresponds for instance to the observations of
\citet{fuhrmann98} for FGK stars in the solar neighborhood. In a
second experiment, we assume a simple linear age versus metallicity
relation with a solar value at present time, and a subsolar [Fe/H] =
$-0.5$ metallicity at 10 Gyr.  Figure~\ref{fg:f_biases2} (bottom
panel) demonstrates that the impact is relatively small on the SFH
with both assumptions. The total age does not strongly depend on the
metallicity, and there is no systematic offset with a linear
metallicity variation as a function of age. This is because the large
age bins are populated with objects having both short and long
main-sequence lifetimes, for high and low mass white dwarfs,
respectively.

We conclude that the two-step feature of the SFH, with an enhanced
formation rate in the last 5~Gyr compared to the range 5 $< {\rm
  Age~(Gyr)} <$ 10, is a significant detection for the 20~pc
sample. The total number of stars formed more than 5~Gyr ago is
30, versus 74 objects at a younger age, which is a 3$\sigma$ result
unlikely to be compromised by biases. On the other hand, smaller scale
fluctuations are unlikely to be significant. The small size of the
sample is the primary uncertainty in the overall derivation of the
SFH, given the error bars in Figure~\ref{fg:f_age}. The most important
additional uncertainty in the lower age bins is the IMF in the solar
neighborhood. Different uncertainties come into play for the oldest
remnants, including the uncertain low-mass end of the initial-final
mass relation, the separation between thin and thick disk populations,
and the bias from the velocity dispersion vs. age
relation. Furthermore, the age of old white dwarfs is uncertain by
$\sim$1~Gyr because of the elusive C/O ratio in the core \citep[see
  Figure 7 of][]{fontaine01}. However, it is fairly evident that the
oldest white dwarfs in the solar neighborhood have an age of at least
8~Gyr due to their unambiguous cool temperatures and cooling times.

\subsection{Comparison with other studies}

\begin{figure}[p]
\epsscale{0.8}
\plotone{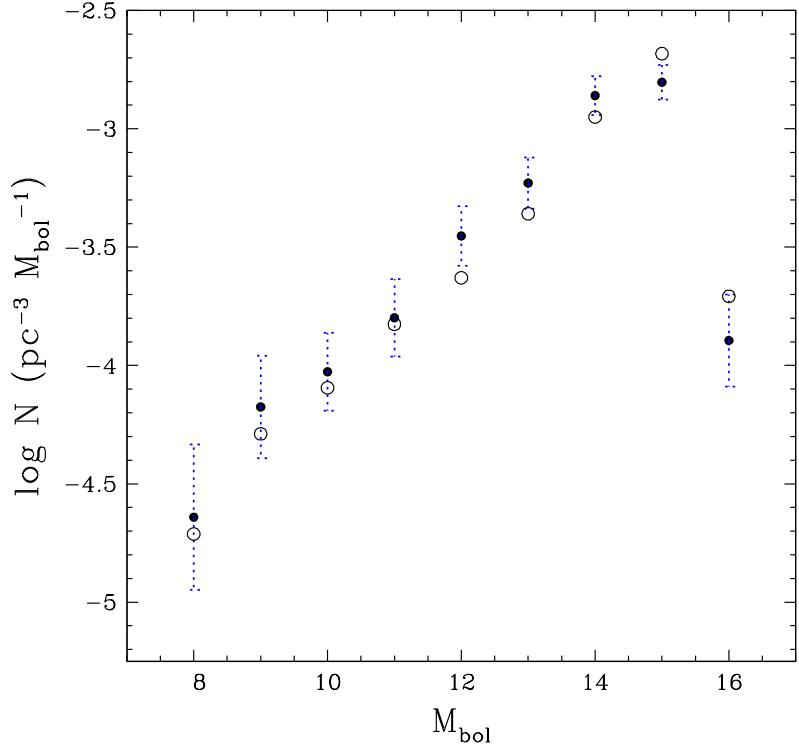}
\begin{flushright}
\caption{Simulated white dwarf luminosity functions for a 100~pc volume
complete sample assuming our derived SFH for the local sample (solid circles),
or a constant SFH for the last 10 Gyr (open circles). In both cases, the 
space density of white dwarfs is fixed at $4.39 \times 10^{-3}$ pc$^{-3}$ (GB12). 
We rely on the main-sequence lifetimes of \citet{hurley00}, the initial-final mass relation of \citet{kalirai08}, the cooling sequences of 
\citet{fontaine01}, the velocity dispersion vs. age relation of Eq.~3 (below 5~Gyr), and suppose a Salpeter IMF. We also present the error bars (blue, dotted) from the number statistics of the local 20~pc sample.
\label{fg:f_lum}}
\end{flushright}
\end{figure}

In principle, our study is most easily compared to the derivation of
the SFH from white dwarf luminosity functions. The main difference
compared to our direct technique is that the information about
individual masses and total ages is lost in the luminosity
function. In order to compare these independent techniques, we
performed Monte Carlo simulations for a volume complete 100~pc sample
using the total space density from GB12. We suppose a Salpeter IMF
while main-sequence lifetimes, cooling sequences, the velocity dispersion
vs. age relation, and the initial-final mass relation are based on the
same models as those described in Section~3. The simulated luminosity
functions presented in Figure~\ref{fg:f_lum} either assume our derived
SFH of Figure~\ref{fg:f_age} (filled points) or a constant SFH in the
last 10 Gyr (open points).  The results suggest that the luminosity
function has a signature of a non-constant SFH. Figure~\ref{fg:f_lum}
also shows the error bars based on number statistics for the 20~pc
sample, illustrating that a larger volume complete sample is necessary
to extract a statistically significant SFH from the luminosity
function.  As a consequence, we make no attempt to model the observed
luminosity function of GB12.  Furthermore, there is no guarantee that
the observed luminosity function defines an unique solution for the SFH.

\citet{SFRWD} derived the SFH from the inversion of the white dwarf
luminosity functions in the SDSS \citep{harris06} and the SuperCOSMOS
Sky Survey \citep{rowell11}. One advantage is that their observed
samples are significantly larger than the local sample. As we did in
this work, they reviewed the different biases impacting their results,
and similarly determined that the initial-final mass relation, IMF,
and initial metallicity have little impact on the SFH. However, they
obtained that the cooling models have a strong impact on the SFH, an
interpretation that we do not support. We suggest instead that the
different cooling sequences highlight a shortcoming in the inversion
technique or the bias corrections for the incompleteness of the
samples. We refrain from a qualitative comparison with the results of
\citet{SFRWD} until these issues are resolved, although we notice that
they observe bimodal distributions for both samples they studied, with
a stellar formation minimum at $\sim$5~Gyr, which is in agreement with
our study. They have significant formation peaks at older ages,
between 7-10~Gyr, which is not supported at face value by our study,
even though our experiment with a high velocity dispersion for old
objects in Figure~\ref{fg:f_biases1} supports a smaller second peak.

\begin{figure}[p]
\epsscale{0.8}
\plotone{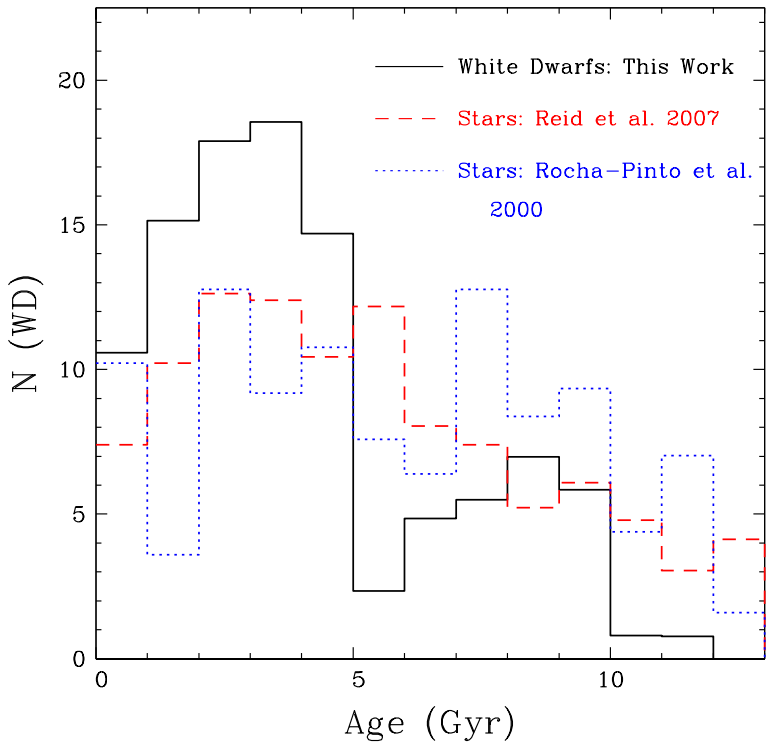}
\begin{flushright}
\caption{Comparison of our derived SFH (solid black) with the studies
  of \citet[][Sample~A, dashed red]{reid07} and \citet[][dotted
    blue]{rocha00} based on the age vs. metallicity distribution of
  \citet{valenti05}, and chromospheric ages, respectively. The stellar
  distributions are re-binned and re-normalized to the white dwarf
  distribution.\label{fg:f_compar1}}
\end{flushright}
\end{figure}

There are many local SFH studies based on stellar observations, using
rather different techniques. None of them are a direct equivalent to
our technique since ages are typically not available for all stars
 in volume-complete samples. \citet{reid07}
relied on a nearly complete sample of $\sim$ 500 stars identified from
Hipparcos with absolute magnitudes $4 < M_{\rm V} < 6$ within 30~pc.
About half of the stars are also part of the \citet{valenti05} sample
of high-resolution echelle spectra from which they determined precise
atmospheric parameters and thereafter derived ages from stellar
isochrones. \citet{reid07} used the age vs. metallicity relation from
this subsample to estimate ages for the other half of the sample,
where metallicities were derived from the Str\"omgren colors.  This
technique is not without problems since there is a significant scatter
in the age vs. metallicity relation and ages well over 15~Gyr are
found for many stars in the sample. In Figure~\ref{fg:f_compar1}, we
compare our results to those of \citet[][Sample~A]{reid07}, showing
that their roughly constant SFH is not compatible with our findings.

\begin{figure}[p]
\epsscale{0.8}
\plotone{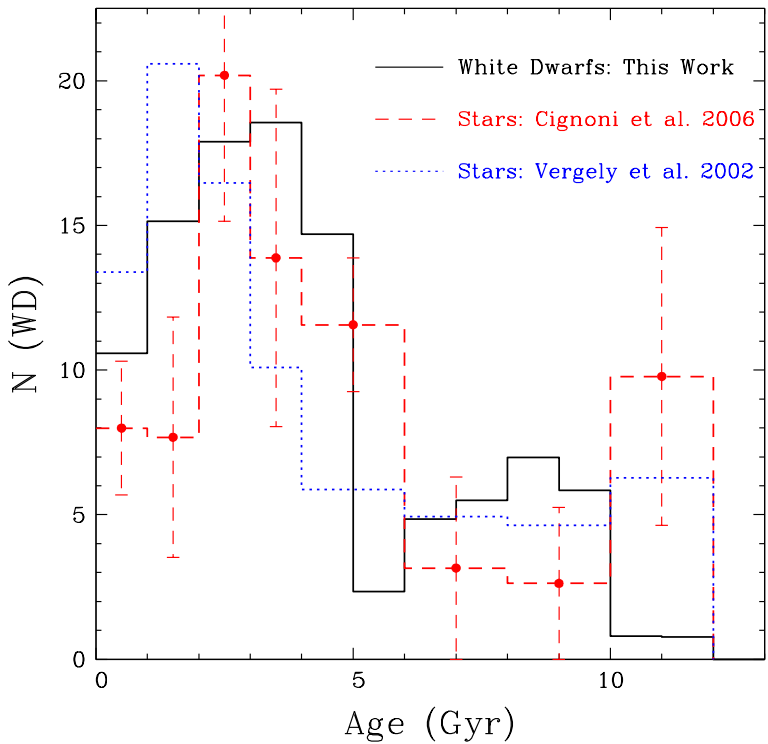}
\begin{flushright}
\caption{Comparison of our derived SFH (solid black) with the studies
  of \citet[][dashed red]{SFR2} and \citet[][dotted blue]{SFR1} based
  on the inversion of the observed Hipparcos color-magnitude diagram
  in the solar neighborhood. The stellar distributions are
  re-normalized to the white dwarf distribution. We also add the
  uncertainties given in \citet{SFR2} as an illustrative example of
  the errors.\label{fg:f_compar2}}
\end{flushright}
\end{figure}

A second group of studies also rely on Hipparcos data of F and G stars
to place them on color-magnitude diagrams. Instead of assigning
individual ages, they rely on Bayesian techniques to invert the
observed color-magnitude diagram into a SFH. Much like the study of
\citet{reid07}, a set of stellar isochrones is at the center of age
determinations. The analysis of \citet{SFR1} constrains simultaneously
the SFH and age vs. metallicity relation, as well as the IMF and SFH
in a separate experiment. Furthermore, they do not limit the distance
in the vertical Galactic coordinate ensuring they are not biased
against velocity dispersion. On the other hand, \citet{SFR2} rely on a
hybrid approach, also inverting the color-magnitude diagram but using
underlying assumptions for the IMF and age vs. metallicity
relation. They rely on a volume complete Hipparcos sample with $M_{\rm
  V} < 3.5$ and distances within 80~pc. They demonstrate that the
recovered SFH is not very sensitive to the assumed IMF, binary
fraction, possible stellar streams, and velocity cuts. However, they
find that the age vs. metallicity relation has a significant impact on
the outcome, and in spite of that, adopt a constant metallicity vs. age relation with a
scatter of $\sigma_{\rm [Fe/H]}$ =~0.2 to match the observations of
\citet{nordstrom04}. The utilization of a different
relation, such as the linear relation employed by \citet{reid07},
would have resulted in a appreciably different SFH (see
Figure~9 of Cignoni et al.). In addition, they suggest that their SFH is significantly
undersampled for ages larger than 7~Gyr due to the lack of good age
tracers with $M_{\rm V} < 3.5$. Figure~\ref{fg:f_compar2} compares our
findings with those of \citet{SFR1} and \citet{SFR2}, showing this
time a relatively good agreement. As an illustration, we also add the
error bars derived by \citet{SFR2}, demonstrating that even though we
have a much smaller number of objects in our sample in
Figure~\ref{fg:f_age}, our error bars are not remarkably
larger. All of the stellar studies described above are also subject to
uncertainties in the predicted isochrones, which are difficult to
constrain and not typically included in the quoted uncertainties
\citep{soderblom10}.

Most other techniques employed to derive the SFH are limited to small
lookback times \citep{fuente04,fuchs09}, but support the view of
\citet{SFR1} and \citet{SFR2} about a formation peak a few Gyr
ago. However, \citet{rocha00} studied the chromospheric age
distribution of 552 late-type dwarfs and converted their results to a
SFH by applying scale height, volume, and stellar evolution
corrections. We present their results in Figure~\ref{fg:f_compar1}
where the SFH is roughly constant with age, once again in opposition
to our results.  While chromospheric age determination is in principle
straightforward and provides ages with a precision of $\sim$0.2 dex
\citep{soderblom91}, the interpretation at young ages differs between
studies \citep{fuchs09} and activity can be difficult to detect in
older stars.

All in all, the comparison of our results with stellar studies is
difficult because of the conflicting derived SFH ranging from constant
values in Figure~\ref{fg:f_compar1} to peaked distributions at young
ages in Figure~\ref{fg:f_compar2}. Our white dwarf derived SFH is in
better overall agreement with the studies of \citet{SFR1} and
\citet{SFR2}, although this does not necessarily support their
techniques over others since the agreement is only qualitative. There
is an agreement over the notable drop in the stellar formation for
ages older than $\sim$5~Gyr and younger than 10~Gyr.  \citet{SFR1} and
\citet{SFR2} predict this transition at younger and older ages that
our study, respectively. For ages larger than 10~Gyr, both stellar
studies predict a significant number of stars although \citet{SFR2}
are cautious and attribute their second peak to undersampling. In our
case, the lack of stars older than 10~Gyr is consistent with the
derived age of the disk from white dwarfs \citep{leggett98}, which is
not entirely surprising since the early studies looking at the age of
remnants used some of the objects and observations from the current local
sample.

Due to the intrinsic brightness of the FGK stars, stellar studies
typically cover a significantly larger volume than our white dwarf sample. However, it does not necessarily imply lower error bars on the SFH since
uncertainties for individual ages are large for stars.  One
advantage of the stellar studies is that they have a better sampling
of the vertical scale height of the disk. Faint stellar streams that
are observed in the solar neighborhood \citep{seabroke07} may impact
the derived SFH, especially for smaller samples, and our derived SFH
within 20~pc may not represent that of the Galactic
disk. Nevertheless, the fact that we recover the results of
\citet{SFR1} and \citet{SFR2} suggests that streams are not an issue.

\section{CONCLUSION}

We presented a new technique where individual white dwarf atmospheric
parameters, for a volume complete sample, are used to derive the
stellar formation history in the solar neighborhood. The method
compares advantageously to other techniques aimed at extracting the
SFH for the Galactic disk. The success of the method resides in the
fact that the white dwarf masses and cooling ages, the main-sequence lifetime as
a function of mass, and the initial-final mass relation are all
relatively well constrained quantities. Therefore, it allows for a
precise transformation of the remnant atmospheric parameters to total
ages and initial masses. The main uncertainties for the age of old
remnants are the scatter in the observed initial-final relation and the
well-known unconstrained composition of the core
\citep{fontaine01}. We found that it is also difficult to identify the
thin or thick disk nature of the old remnants, although it does not
impact significantly our derivation of the overall SFH.  At very young
ages, the main shortcoming is the lack of stars that became white
dwarfs.  Finally, the largest limitation of the current analysis is
the small size of the 20~pc sample.  However, future surveys like {\it
  Gaia} will resolve this issue.

The SFH derived from white dwarfs was compared to similar studies
relying on large samples of FGK stars and the chromospheric activity
in late-type dwarfs.  There are conflicting results in these studies,
and we suggest that the SFH from white dwarfs may be the most accurate at
intermediate and large ages. We recover a significant enhanced
formation rate in the last $\sim$5 Gyr by a factor $\sim$2.5 of
compared to the range 5-10 Gyr. This result is in agreement with a
number of studies looking at the stellar content of the solar
neighborhood.

\acknowledgements

Support for this work was provided by NASA through Hubble Fellowship
grant \#HF-51329.01 awarded by the Space Telescope Science Institute,
which is operated by the Association of Universities for Research in
Astronomy, Inc., for NASA, under contract NAS 5-26555. This project
was supported by the National Science Foundation (NSF) through grant
AST-1211719.

\clearpage

\end{document}